\documentclass[showpacs,fleqn,nobibnotes]{revtex4}

\usepackage{amsmath}
\usepackage{graphicx}

\def\lsim{\raise0.3ex\hbox{$<$\kern-0.75em\raise-1.1ex\hbox{$\sim$}}}
\def\gsim{\raise0.3ex\hbox{$>$\kern-0.75em\raise-1.1ex\hbox{$\sim$}}}

\newcommand{\rr}{\mbox{\boldmath $r$}}

\newcommand{\rb}{\mbox{\boldmath $b$}}

\newcommand{\rd}{\mbox{\boldmath $\Delta$}}

\newcommand{\aaa}{\mathcal{A}}

\newcommand{\N}{\mathcal{N}}

\begin{document}

\title{Diffractive $\rho$ production  at small $x$ in  future Electron - Ion Colliders}
\author{V. P. Gon\c{c}alves $^{1,2}$, F. S. Navarra $^{3}$  and D. Spiering $^{3}$}

\affiliation{ $^{1}$ Department of Astronomy and Theoretical Physics, Lund University, 223-62 Lund, Sweden.}
\affiliation{$^{2}$ Instituto de F\'{\i}sica e Matem\'atica,  Universidade
Federal de Pelotas, 
Caixa Postal 354, CEP 96010-900, Pelotas, RS, Brazil}
\affiliation{$^{3}$ Instituto de F\'{\i}sica, Universidade de S\~{a}o Paulo, CEP 05315-970 S\~{a}o Paulo, SP, Brazil.}

\begin{abstract}
The future Electron - Ion ($eA$) Collider is expected to probe the high energy regime of the QCD dynamics, with the exclusive vector meson production cross section being one of the most promising observables. In this paper we complement previous studies of   exclusive processes  presenting a comprehensive analysis of  diffractive $\rho$ production at small  $x$. We compute the coherent and incoherent cross sections  taking into account  non-linear QCD dynamical effects and considering different models for the dipole - proton scattering amplitude and for the vector meson wave function. The dependence of  these cross sections with the energy, photon virtuality, nuclear mass number 
and squared momentum transfer is analysed in detail. Moreover, we compare the non-linear predictions with those obtained in the linear regime. Finally, we also estimate the exclusive photon, $J/\Psi$ and $\phi$ production and compare with  the results obtained for  $\rho$ production. Our results demonstrate that the analysis of  diffractive $\rho$ production in  future electron - ion colliders will be important  to understand the non-linear QCD dynamics.  
\end{abstract}

\maketitle

\section{Introduction}

One of the main goals of the future Electron - Ion Colliders (EIC) \cite{Raju,Boer,Accardi,LHeC} is the study of the hadronic structure in the non-linear regime of the Quantum Chromodynamics (QCD). Theoretically,  the magnitude of the non-linear 
effects is expected to be amplified by the nuclear medium, which should allow us  to determine the presence of gluon saturation effects, their  magnitude  and 
which is the correct theoretical framework for their description (For reviews see, e.g., Ref. \cite{hdqcd}).
During the last years  many authors have proposed the study of several inclusive and diffractive observables 
\cite{victor,simone1,simone2,Nik_schafer,Kowalski_prl,Kowalski_prc,erike_ea1,erike_ea2,vmprc,simone_hq,
Caldwell,vic_erike,Lappi_inc,Toll,joao,Lappi_bal,diego} in order to search for non-linear effects in $eA$ collisions. One of the most promising 
observables is the  exclusive production of vector mesons or photons, which are experimentally clean and can be unambiguously identified by the 
presence of a rapidity gap in the final state. As these processes are driven by the gluon  content of the target, with the  cross sections being  
proportional to the square of the scattering amplitude,  they  are strongly sensitive to the underlying QCD dynamics. The diffractive production 
of $J/\Psi$ and $\phi$  in $eA$ processes was analysed in detail in Refs.  \cite{vmprc,vic_erike,Caldwell,Lappi_inc,Toll} taking into account   
non-linear effects, and predictions for the energy, virtuality and transverse momentum dependencies were presented. In contrast, in the case of 
$\rho$ production, only predictions for the energy and virtuality dependencies were presented in Refs. \cite{vmprc,vic_erike}. Our goal in this 
paper is to complement and extend these previous studies and present a comprehensive analysis of  $\rho$ production. We present, for the first 
time,  predictions for the squared  momentum transfer ($t$) distributions, which are an important source of information about the 
spatial distribution of the gluons in a nucleus and about fluctuations  of the nuclear color fields. We take into account  non-linear QCD effects and consider different models for the dipole - proton scattering amplitude as well as for the vector meson wave function. Moreover, we present 
a comparison between the non-linear predictions with those obtained using a linear model for the QCD dynamics. Finally, we present a comparison between our predictions for the $\rho$ and those for the exclusive photon, $J/\Psi$ and $\phi$ production.

This paper is organized as follows. 
In the next Section we present a brief review of the description of  diffractive $\rho$ production  in the color dipole formalism. 
In Section \ref{res} we present our results for the energy, virtuality and momentum transfer dependencies of the $\rho$ cross section.  A comparison between  our  predictions for the $\rho$ production with those for other final states and the impact of the non-linear effects in the $t$ - distributions is investigated. Finally, in Section \ref{conc} we summarize our main conclusions.

\section{Diffractive $\rho$ production in the color dipole picture}
\label{nucrho}

\begin{figure}[t]
\begin{center}
\scalebox{0.35}{\includegraphics{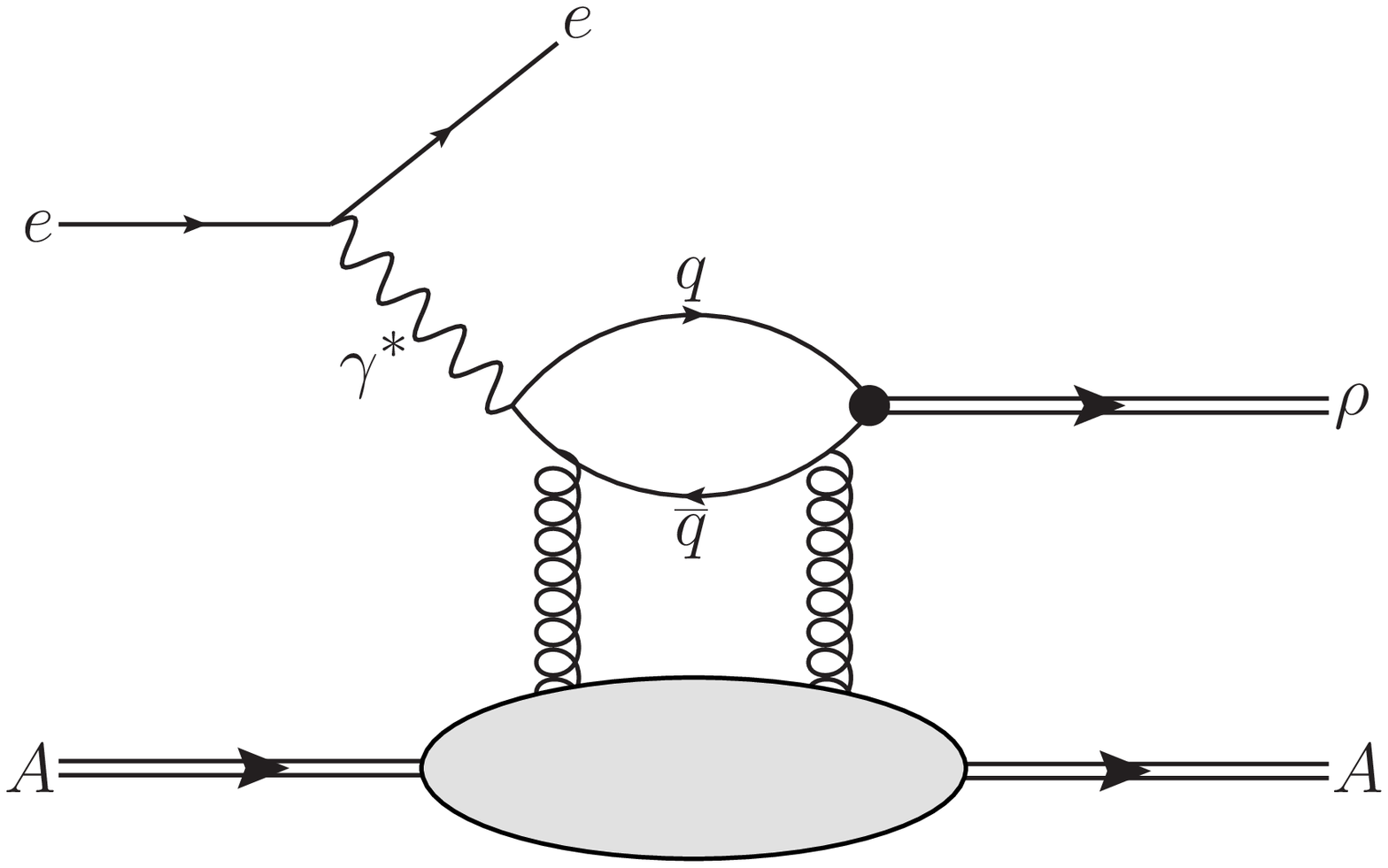}}
\scalebox{0.35}{\includegraphics{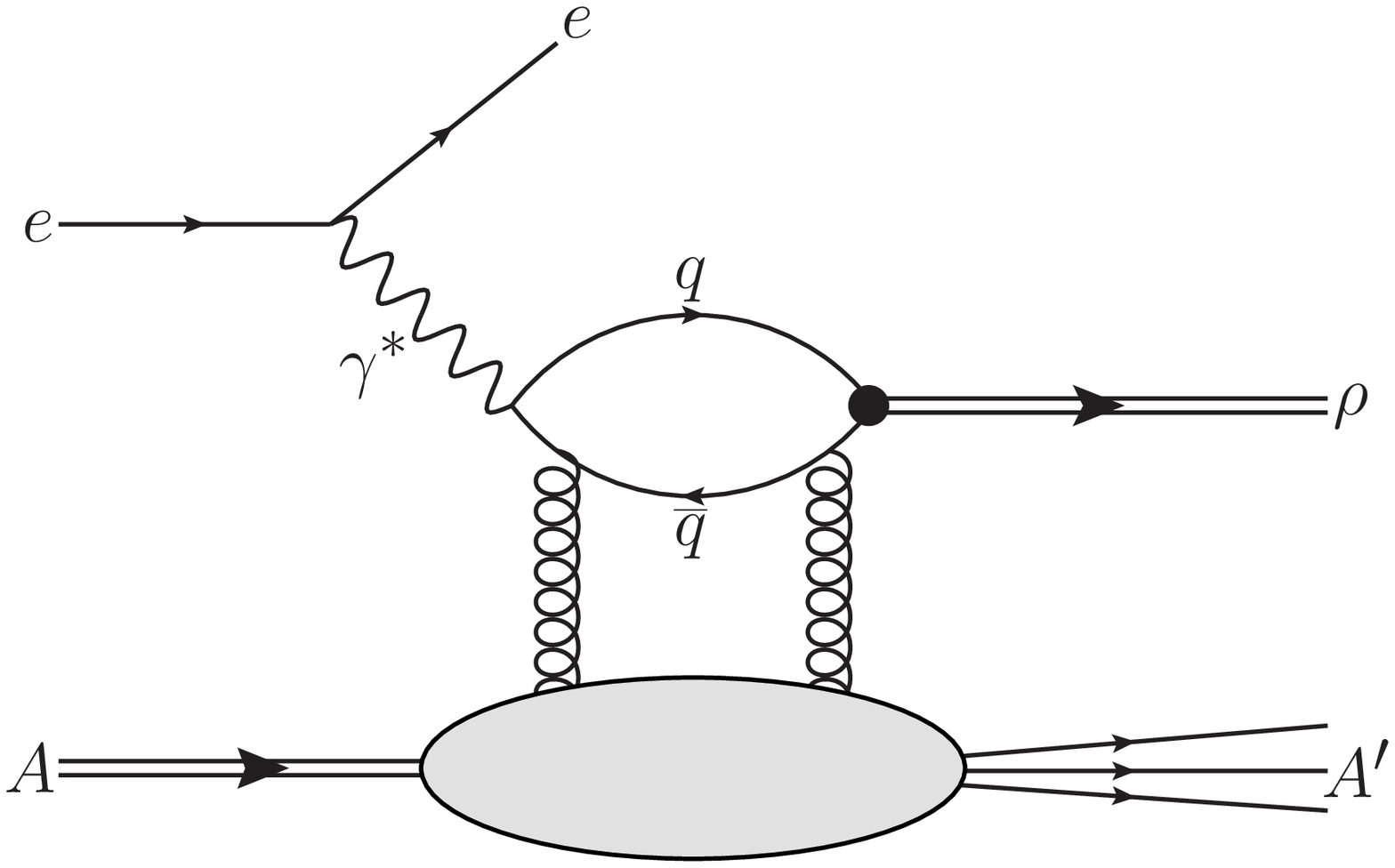}}
\caption{Diffractive $\rho$ production in  coherent (left panel) and  incoherent (right panel) $eA$ collisions.}
\label{diagramas}
\end{center}
\end{figure}

In the color dipole picture,  the  $eA \rightarrow e\rho Y$ process can be factorized in terms of the fluctuation of the virtual photon into a $q \bar{q}$ color dipole, the dipole-nucleus scattering by a color singlet exchange  and the recombination into the exclusive final state $\rho$. The final state is characterized by the presence of a rapidity gap. The dipole - nucleus interactions can be classified as coherent or incoherent. If the nucleus scatters elastically, $Y = A$, the process is called coherent production  and can be represented by the diagram in Fig. \ref{diagramas} (left panel). On the other hand, if the nucleus scatters inelastically, i.e. breaks up ($Y = A^{\prime}$),   the process is denoted incoherent production and can be represented as in Fig. \ref{diagramas} (right panel). As discussed e.g. in Refs. \cite{Caldwell,Lappi_inc,Toll}, these different processes probe distinct properties of the gluon density of the nucleus. While coherent processes  probe the  average spatial distribution of gluons, the incoherent ones are determined by fluctuations and correlations in the gluon density. In order to access these informations, it is fundamental to   understand the contribution of each process in distinct kinematical regimes. One of our main motivations is to study the energy, virtuality and transverse momentum dependencies of the $\rho$ cross section. Another one, is the fact that the study of the $\rho$ production at different photon virtualities allows to probe the transition between the non-linear and linear regimes of the QCD dynamics. While at large - $Q^2$ the process is dominated by small size dipoles, where the contribution of non-linear effects is negligible, at small photon virtualities the main contribution comes from  large size dipoles, with the dynamics being determined by the gluon saturation effects. In contrast,  $J/\Psi$ production is dominated by small dipoles, which implies a smaller contribution of  non-linear effects. Therefore, the study of $\rho$ production in coherent and incoherent processes is ideal to investigate in detail the impact of the saturation effects in the spatial distribution, fluctuations and correlations in the nuclear gluon density.  

\subsection{$ep$ collisions}

Let us start our analysis presenting a brief review of the description of the diffractive $\rho$ production in $ep$ collisions at high energies.
The $\gamma^* p \rightarrow \rho p$  cross section can be written as
\begin{eqnarray}
\sigma (\gamma^* p \rightarrow \rho p) =  \int_{-\infty}^0 \frac{d\sigma}{d{t}}\, d{t}  
= \frac{1}{16\pi}  \sum_{i=T,L}\int_{-\infty}^0 |{\cal{A}}_i^{\gamma^* p \rightarrow \rho p }(x, Q^2, \Delta)|^2 \, d{t}\,\,,
\label{sctotal_intt}
\end{eqnarray}
with the  amplitude for producing an exclusive $\rho$ meson diffractively in an electron - proton scattering being given by
\begin{eqnarray}
 {\cal A}_{T,L}^{\gamma^* p \rightarrow \rho p}({x},Q^2,\Delta)  =  i
\int dz \, d^2\rr \, d^2\rb_p  e^{-i[\rb_p-(1-z)\rr].\rd} 
 \,\, (\Psi^{\rho*}\Psi)_{T,L}  \,\,2 {\cal{N}}^p({x},\rr,\rb_p)
\label{amp}
\end{eqnarray}
where $T$ and $L$ denotes the transverse and longitudinal polarizations of the virtual photon, $(\Psi^{\rho*}\Psi)_{i}$ denotes the wave function overlap between the virtual photon and $\rho$ meson wave functions, $\Delta = - \sqrt{t}$ is the momentum transfer and $\rb_p$ is the impact parameter of the dipole relative to the proton. Moreover, the variables  $\rr$ and $z$ are the dipole transverse radius and the momentum fraction of the photon carried by a quark (an antiquark carries then $1-z$), respectively. $ {\cal N}^p (x, \rr, \rb_p)$ is the forward dipole-proton scattering amplitude (for a dipole at  impact parameter $\rb_p$) which encodes all the information about the hadronic scattering, and thus about the non-linear and quantum effects in the hadron wave function.
It  depends on the $\gamma$ -- hadron center - of - mass reaction energy, $W$, through the variable:
\begin{equation}
x = \frac{ Q^2 + m^2_V}{Q^2 + W^2}
\label{defex}
\end{equation}
where $Q^2$ and $m_V$ are the photon virtuality and vector meson mass, respectively. Consequently,  the cross section for $\rho$ 
production at low $Q^2$ is much more sensitive to low $x$ effects than the  one for  production of heavy mesons.

At high energies the evolution with the rapidity $Y = \ln \frac{1}{x}$ of
 $\mathcal{N}^p(x,\rr,\rb_p)$  is given in the Color Glass Condensate (CGC) formalism \cite{CGC} by the infinite hierarchy of equations, the so 
called
Balitsky-JIMWLK equations \cite{BAL,CGC}, which reduces in the mean field approximation to the Balitsky-Kovchegov (BK) equation \cite{BAL,kov}. This equation has been solved in Ref. \cite{bkrunning} taking into account the running coupling corrections
 and assuming 
the translational invariance approximation, which implies  $\mathcal{N}^p(x,\rr,\rb_p) = \mathcal{N}^p(x,\rr) S(\rb_p)$. As a consequence, the dipole - proton cross section, defined by 
\begin{equation} 
\sigma_{dp} (x, \rr) = 2 \int d^2 \rb_p \,  {\cal N}^p (x, \rr, \rb_p)\,\,,
\label{sdip}
\end{equation}
will be given by $\sigma_{dp} (x, \rr) = \sigma_0 \cdot  {\cal N}^p (x, \rr)$, with the normalization of the dipole cross section ($\sigma_0$) being fitted to data. Hereafter we will denote by rcBK the predictions obtained using the numerical solution of  running coupling BK equation for ${\cal N}^p (x, \rr)$ obtained in Ref. \cite{bkrunning}. Moreover, we will also consider  the bCGC model, which is based on the CGC formalism and takes into account the impact parameter dependence 
of the dipole - proton scattering amplitude.   In this model the dipole - proton scattering amplitude is given by \cite{kmw} 
\begin{widetext}
\begin{eqnarray}
\mathcal{N}^p(x,\rr,\rb_p) =   
\left\{ \begin{array}{ll} 
{\mathcal N}_0\, \left(\frac{ r \, Q_s(b_p)}{2}\right)^{2\left(\gamma_s + 
\frac{\ln (2/r Q_s(b_p))}{\kappa \,\lambda \,Y}\right)}  & \mbox{$r Q_s(b_p) \le 2$} \\
 1 - e^{-A\,\ln^2\,(B \, r  Q_s(b_p))}   & \mbox{$r Q_s(b_p)  > 2$} 
\end{array} \right.
\label{eq:bcgc}
\end{eqnarray}
\end{widetext} 
with  $\kappa = \chi''(\gamma_s)/\chi'(\gamma_s)$, where $\chi$ is the 
LO BFKL characteristic function.  The coefficients $A$ and $B$  
are determined uniquely from the condition that $\mathcal{N}^p(x,\rr,\rb_p)$, and its derivative 
with respect to $r\,Q_s(b_p)$, are continuous at $r\,Q_s(b_p)=2$. The impact parameter dependence of the  proton saturation scale $Q_s(b_p)$  is given by:
\begin{equation} 
  Q_s(b_p)\equiv Q_s(x,b_p)=\left(\frac{x_0}{x}\right)^{\frac{\lambda}{2}}\;
\left[\exp\left(-\frac{{b_p}^2}{2B_{\rm CGC}}\right)\right]^{\frac{1}{2\gamma_s}},
\label{newqs}
\end{equation}
with the parameter $B_{\rm CGC}$  being obtained by a fit of the $t$-dependence of exclusive $J/\psi$ photoproduction.  
Moreover, the factors $\mathcal{N}_0$ and  $\gamma_s$  were  taken  to be free. In what follows we consider the set of parameter obtained in Ref. \cite{amir} by fitting the recent HERA data for the reduced $ep$ cross sections:
 $\gamma_s = 0.6599$, $\kappa = 9.9$, $B_{CGC} = 5.5$ GeV$^{-2}$, $\mathcal{N}_0 = 0.3358$, $x_0 = 0.00105$ and $\lambda = 0.2063$.  In our calculations we will consider the rcBK and bCGC models in order to estimate  the dependence of our predictions on the model used to describe the dipole - proton amplitude. It is important to emphasize that both models predict the saturation of ${\cal N}^p$ at large dipole sizes and that the non-linear effects become negligible at small - $\rr$. However, these models predict a different behavior in the transition region between small and large dipoles, as well as distinct energy dependence for the saturation scale.

The precise form of the wave function overlap $(\Psi^{\rho*}\Psi)_{i}$ in Eq. [\ref{amp}]  is still  an open question. In contrast to the photon wave function, which is well known in the literature (See e.g. \cite{kmw}),  the $\rho$ wave 
function is still a subject of investigation.  The simplest approach  is to assume that the $\rho$ is predominantly a quark-antiquark state 
and that the spin and polarization structure is the same as in the  photon \cite{dgkp,nnpz,sandapen,KT} (for other approaches see, for example, 
Ref. \cite{pacheco}). As a consequence, the wave function overlap is given by (For details see Ref. \cite{kmw})
\begin{eqnarray}
  (\Psi^*_V\Psi)_T &=& \frac{\hat e_fe}{4\pi}\frac{N_c}{\pi z(1-z)}
    \left\{m_f^2K_0(\epsilon r)\phi_T(r,z)-\left[z^2+(1-z)^2\right]\epsilon K_1(\epsilon r)\partial_r\phi_T(r,z)\right\}, \\
  (\Psi^*_V\Psi)_L &=& \frac{\hat e_fe}{4\pi}\frac{N_c}{\pi}2Qz(1-z)K_0(\epsilon r) 
    \left[M_V\phi_L(r,z) + \delta \frac{m_f^2-\nabla_r^2}{M_Vz(1-z)} \phi_L(r,z) \right],
\end{eqnarray}
where $ \hat{e}_f $ is the effective charge of the vector meson, $m_f$ is the quark mass, $N_c = 3$, $\epsilon^2 = z(1-z)Q^2 + m_f^2$   and $\phi_i(r,z)$ define the scalar part of the  vector meson wave functions. In order to estimate the dependence of our predictions on the model for the $\rho$ wave function, in what follows we will consider  
the Boosted Gaussian ($\delta=1$) and Gaus-LC ($\delta=0$) models  for $\phi_T(r,z)$ and $\phi_L(r,z)$, which are largely used in the literature.  
In the Boosted Gaussian model the functions $\phi_i(r,z)$ are given by \cite{kmw}
\begin{eqnarray}
  \phi_{T,L}(r,z) =  N_{T,L}\,z(1-z)\exp
  \left[ -\frac{m_f^2 R^2}{8z(1-z)} - \frac{2z(1-z)r^2}{ R^2} + \frac{m_f^2 {R}^2}{2} \right].
\end{eqnarray}
 In contrast, in the Gaus-LC model, they are given by \cite{kmw}
\begin{eqnarray}
  \phi_T(r,z) &=& N_T\left[z(1-z)\right]^2\exp\left(-r^2/2R_T^2\right),\\
  \phi_L(r,z) &=& N_L z(1-z)\exp\left(-r^2/2R_L^2\right).
\end{eqnarray}
The parameters $N_i$, $R$ and $R_i$ are  determined by the normalization condition of the wave function and by the decay width. 
In Table I  we present the value of these parameters for the $\rho$ wave function.
As demonstrated in Ref. \cite{kmw}, the dipole size dependence of the  overlap functions are very similar for the longitudinal polarization. On the other hand,  in the case of transverse photons, the predictions differ at small values of $Q^2$.
The impact of using  different models in the  $ep$ cross section is small,  with the  predictions being compatible with HERA data within the experimental errors (See e.g. Refs. \cite{kmw,amir,anelise}). Similar conclusions have been obtained in  the analysis of the vector meson production in ultraperipheral collisions performed in Ref. \cite{vicper}.  However, the impact on the coherent and incoherent $eA$ cross sections is still  an open question, which should be quantified in order to obtain realistic predictions. This is one of the main motivations to present the Boosted Gaussian and Gauss - LC predictions in the next section.

\begin{table}[t] 
\centering
\begin{tabular}{cccccccc} 
\hline 
Model & $M_{\rho}/\mbox{GeV}$ & $m_{f}/\mbox{GeV}$ & $N_{T}$ & $R_T^{2}/\mbox{GeV}^{-2}$ & $R^{2}/\mbox{GeV}^{-2}$ & $N_{L}$ & 
$R_L^{2}/\mbox{GeV}^{-2}$   \\ 
\hline
\hline
Gaus-LC & 0.776 & 0.14 & 4.47  & 21.9 & -- & 1.79 & 10.4 \\
Boosted Gaussian & 0.776 & 0.14 & 0.911 & -- & 12.9 & 0.853 & -- \\
\hline
\label{tab1}
\end{tabular}
\caption{Parameters of the Gaus-LC and Boosted Gaussian models for the $\rho$ wave function.} 
\end{table}

\subsection{$eA$ collisions}

Let us discuss now diffractive $\rho$ production in $eA$ collisions. In this case we must consider separately the coherent and incoherent processes. In the coherent case, the nucleus is required to remain in its ground state, i.e. intact after the interaction. It corresponds to take the average over the nuclear wave function at the level of the scattering amplitude. In other words, the coherent cross section is obtained by averaging the amplitude before squaring it. Consequently, the differential distribution (with respect to the squared momentum transfer $t$) for coherent interactions will be given by
\begin{eqnarray}
 \frac{d\sigma_{coh}}{dt} = \frac{1}{16\pi}\left|\langle \aaa^{\gamma^*A\rightarrow\rho A}(x,Q^2,\Delta_A) \rangle \right|^2,
\end{eqnarray}
where 
\begin{eqnarray}
 \langle \aaa^{\gamma^*A\rightarrow\rho A}(x,Q^2,\Delta_A) \rangle = i \int d^2 \rb_A  
 \int dz \, d^2\rr   \,  e^{-i\rb_A\cdot\Delta_A}  \,\, (\Psi^{\rho*}\Psi)_{T,L}  \,\,2 {\cal{N}}^A({x},\rr,\rb_A)  ,
\end{eqnarray}
with $\Delta_A^2 =-t$ and $\rb_A$ being the impact parameter between the dipole and the nucleus. At large nuclei, the forward dipole-nucleus amplitude can be expressed as follows
\begin{eqnarray}
{\cal{N}}^A(x,\rr,\rb_A) = 1 - \exp \left[-\frac{1}{2}  \, \sigma_{dp}(x,\rr^2) 
\,A\,T_A(\rb_A)\right] \,\,,
\label{enenuc}
\end{eqnarray}
where $\sigma_{dp}$ is the dipole-proton cross section and $T_A(\rb_A)$ is  the nuclear profile 
function, which is obtained from a 3-parameter Fermi distribution for the nuclear
density normalized to $1$. The above expression can be derived considering the  Glauber-Gribov formalism \cite{gribov}, and it takes into  account the 
multiple elastic rescattering diagrams of the $q \overline{q}$ pair. It is justified  in the large coherence length regime ($l_c \gg R_A$), where the transverse separation $\rr$ of partons 
in the multiparton Fock state of the photon becomes a conserved quantity, {\it i.e.} 
the size 
of the pair $\rr$ becomes eigenvalue
of the scattering matrix. It is important to emphasize that this model for ${\cal{N}}^A$  allows to describe the current  experimental data on the nuclear 
structure function \cite{armesto,erike_ea1,erike_ea2} and it has been used in previous studies \cite{vmprc,vic_erike,diego}  to estimate inclusive and exclusive $eA$ observables.
The corresponding integrated cross 
section  will be given by
\cite{vmprc,kop1}
\begin{eqnarray}
\sigma^{coh}\, (\gamma^* A \rightarrow \rho A)  =  \int d^2\rb_A \left| \int d^2\rr
 \int dz  \,  \Psi_{\rho}^*(\rr,z) \, \mathcal{N}^A(x,\rr,\rb_A)\, \Psi_{\gamma^*}(\rr,z,Q^2)  \right|^2 \,\,.
\label{totalcscoe}
\end{eqnarray}
On the other hand, if the average over the nucleon positions is at the cross section level, the nucleus can break up and the resulting incoherent cross section will be proportional to the variance of the amplitude with respect to the nucleon configurations of the nucleus, i.e., it will measure the fluctuations of the gluon density inside the nucleus (For a detailed derivation see e.g. Refs. \cite{Lappi_inc, Toll}). The integrated cross section for incoherent processes can be expressed as \cite{vmprc,kop1}
\begin{eqnarray}
\sigma^{inc}\, (\gamma^* A \rightarrow \rho A^{\prime}) = \frac{|{\cal I}m \, 
{\cal A}(s,\,t=0)|^2}{16\pi\,B} \;
\label{totalcsinc}
\end{eqnarray}
where at high energies ($l_c \gg R_A$) \cite{kop1}:
\begin{eqnarray}
|{\cal I}m \, {\cal A}|^2  =  \int d^2\rb_A \, A\,T_A(\rb_A)
\left|
\int d^2\rr
 \int dz  \,  \Psi_{\rho}^*(\rr,z) \,\left[\sigma_{dp} \, 
\exp\left(- \frac{1}{2} \, \sigma_{dp} \, A\,T_A(\rb_A)\right)\right]  \Psi_{\gamma^*}(\rr,z,Q^2)  \right|^2 
\label{totalcsinc1}
\end{eqnarray}
with the  $t$  slope ($B$) being  the same as in the case of a nucleon target. 
In the incoherent case, the $q\bar{q}$ pair attenuates with a constant absorption cross 
section, as in the Glauber model, except that the whole exponential is averaged 
rather than just the cross section in the exponent.  For the calculation of the differential cross section $d\sigma/dt$ 
for incoherent interactions we apply for the $\rho$ production the treatment presented in Ref. \cite{Lappi_inc}, which is valid for $t\neq 0$. Consequently, 
we have that 
\begin{eqnarray}
 \frac{d\sigma_{inc}}{dt} = \frac{1}{16\pi} \sum_{i= T,L} \int dz dz^{\prime} d^2\rr d^2\rr^{\prime}
 (\Psi^{\rho*}\Psi)_{i}(z,\rr,Q^2) (\Psi^{\rho*}\Psi)_{i}(z^{\prime},\rr^{\prime},Q^2) \, \langle |{\cal{A}}|^2\rangle \,\,,
 \label{difinc}
\end{eqnarray} 
 with the average of the squared scattering amplitude  being  approximated by \cite{Lappi_inc}
\begin{widetext}
\begin{eqnarray}
\langle |{\cal{A}}(\rr,\rr^{\prime},t)|^2\rangle = 16\pi^2 B_p^2 &\displaystyle\int& d^2\rb 
e^{-B_p\Delta_A^2}\N^p(x,\rr)\N^p(x,\rr^{\prime})\,A\,T_A(\rb_A) \nonumber\\
  &\times&\exp\Big\{ -2\pi(A-1)B_pT_A(\rb_A)\big[\N^p(x,\rr)+\N^p(x,\rr^{\prime})\big] \Big\},
  \label{amp_inc}
\end{eqnarray}
\end{widetext}
where  $\N^p(x,\rr)$ is the dipole - proton scattering amplitude. The parameter $B_p$ is associated to the impact parameter profile function in the proton and can be determined by the normalization $\sigma_0$ of the dipole - proton cross section, since $B_p = \sigma_0 / 4 \pi$.

\begin{figure}[t]
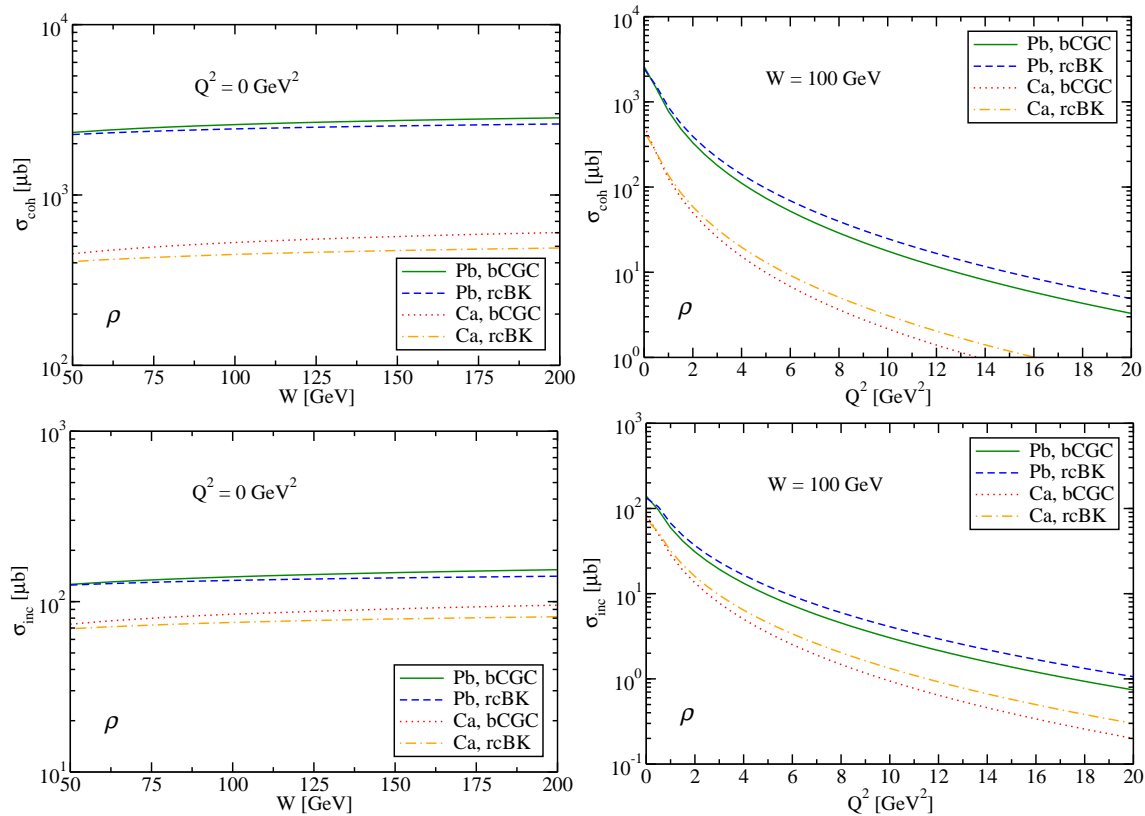

\begin{center}
\scalebox{0.3}{\includegraphics{coherent_q0.eps}}
\scalebox{0.3}{\includegraphics{coherent_w100.eps}} \\
\scalebox{0.3}{\includegraphics{incoherent_q0.eps}}
\scalebox{0.3}{\includegraphics{incoherent_w100.eps}}
\caption{(Color online) Energy and virtuality dependencies of the coherent (upper panels) and incoherent (lower panels) cross sections for two different nuclei. Predictions obtained  using the Gaus-LC model for the $\rho$ wave function.}
\label{sec_energy}
\end{center}
\end{figure}

\begin{figure}[t]
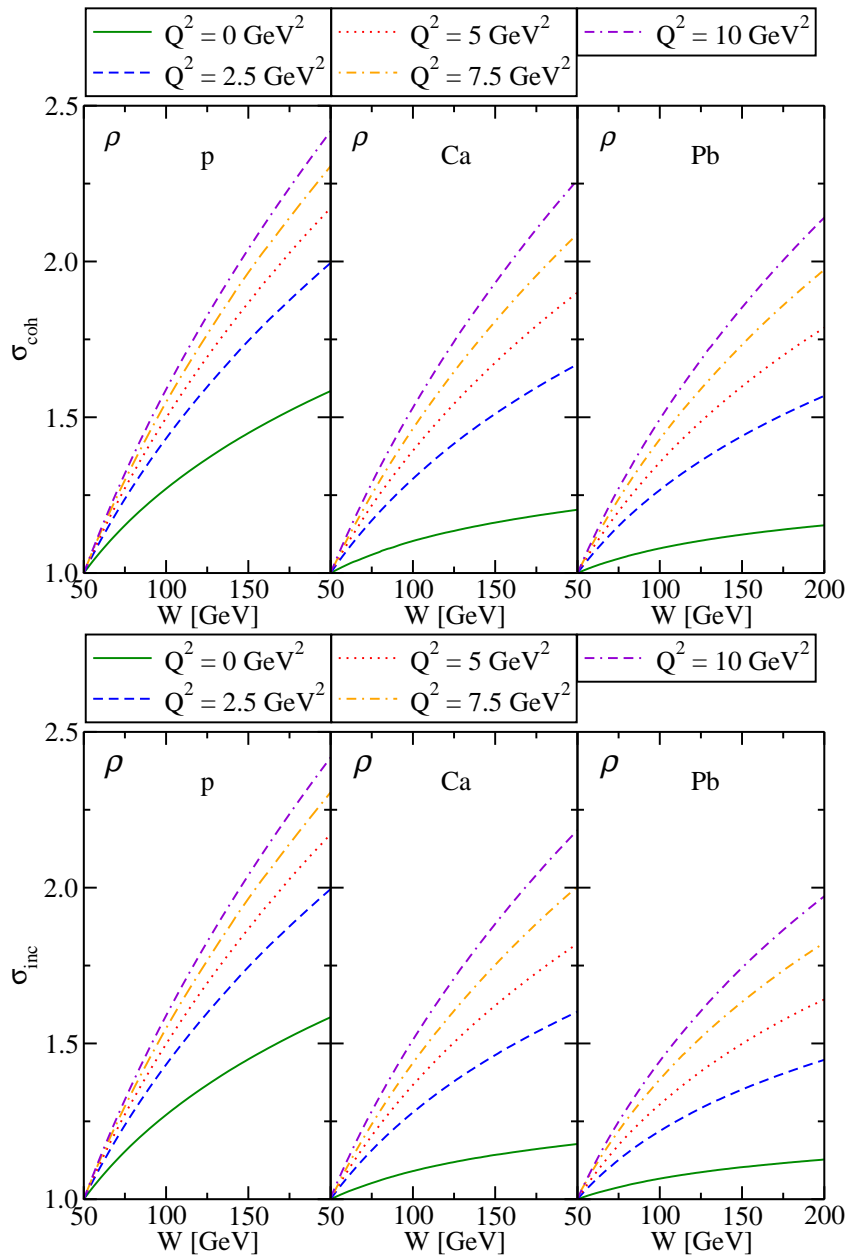

\begin{center}
\scalebox{0.4}{\includegraphics{slopes_coh.eps}}\\
\scalebox{0.4}{\includegraphics{slopes_inc.eps}}
\caption{(Color online) Energy dependence of the normalized coherent (upper panels) and incoherent (lower panels) cross sections for different photon virtualities $Q^2$. The results for the $\rho$ production in $ep$ collisions are presented for comparison. Predictions obtained  using the Gaus-LC model for the $\rho$ wave function and the bCGC model for the dipole - proton scattering amplitude.}
\label{slopes}
\end{center}
\end{figure}

\begin{figure}[t]
\begin{center}
\scalebox{0.5}{\includegraphics{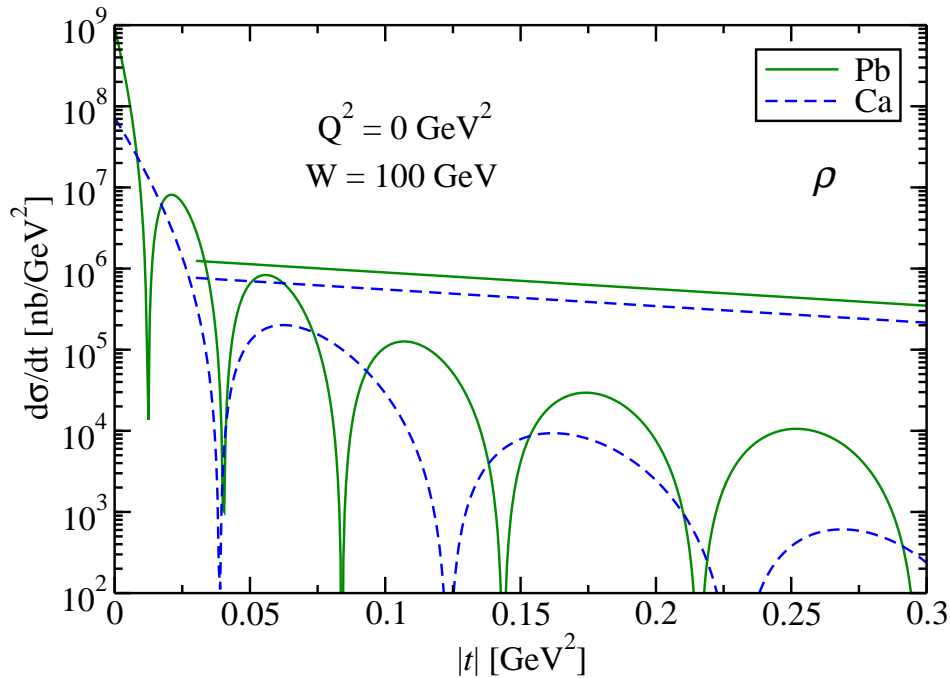}}
\caption{(Color online) Differential cross section for coherent and incoherent interactions for different nuclei. Predictions obtained  using the Gaus-LC model for the $\rho$ wave function and the bCGC model for the dipole - proton scattering amplitude.}
\label{dsdt0}
\end{center}
\end{figure}

\begin{figure}[t]
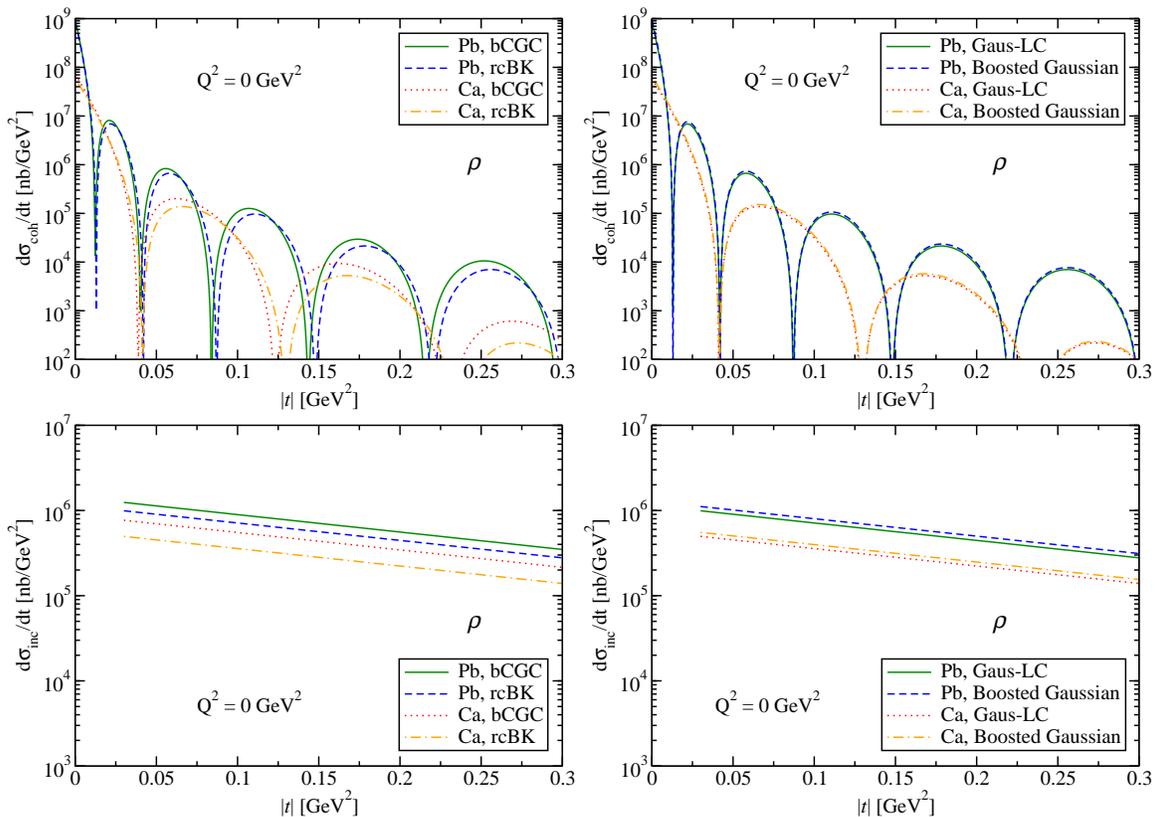

\begin{center}
 \includegraphics[scale=0.3]{dsdt_coherent_q0.eps}
 \includegraphics[scale=0.3]{dsdt_coherent_wave_q0.eps} \\
  \includegraphics[scale=0.3]{dsdt_incoherent_q0.eps}
 \includegraphics[scale=0.3]{dsdt_incoherent_wave_q0.eps}
\caption{(Color online) Differential cross section for coherent (upper panels) and incoherent (lower panels) interactions. Left panels:  Predictions obtained considering  different models for the dipole - proton scattering amplitude  and the Gaus-LC model for the $\rho$ wave function. Right panels:  Predictions obtained considering  different models for the  $\rho$ wave function and the bCGC model for the dipole - proton scattering amplitude.}
\label{dsdt1}
\end{center}
\end{figure}

\begin{figure}[t]
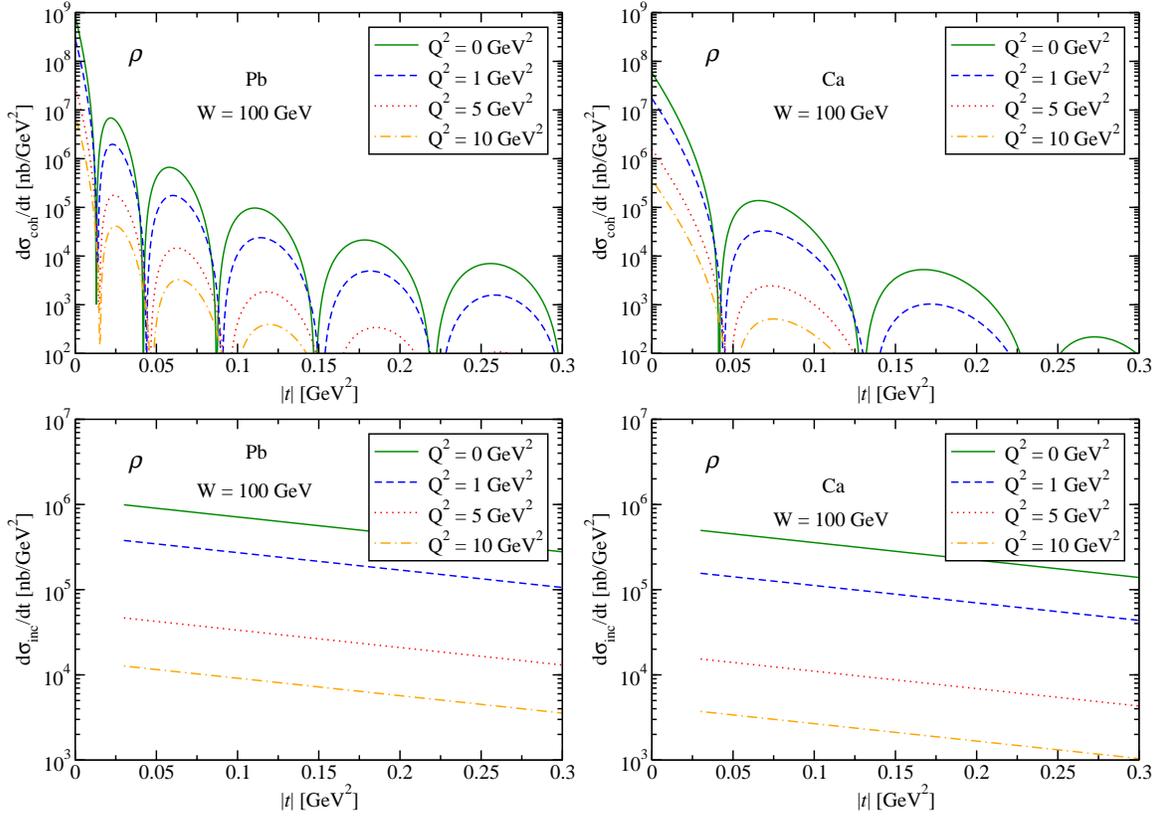

\begin{center}
 \includegraphics[scale=0.3]{dsdt_coherent_pb_q2.eps}
 \includegraphics[scale=0.3]{dsdt_coherent_ca_q2.eps} \\
  \includegraphics[scale=0.3]{dsdt_incoherent_pb_q2.eps}
 \includegraphics[scale=0.3]{dsdt_incoherent_ca_q2.eps}
\caption{(Color online)  Differential cross section for coherent (upper panels) and incoherent (lower panels) interactions considering 
different values of $Q^2$ and distinct nuclei. Predictions obtained using the  Gaus-LC and bCGC models.}
\label{dsdt2}
\end{center}
\end{figure}

\begin{figure}[t]
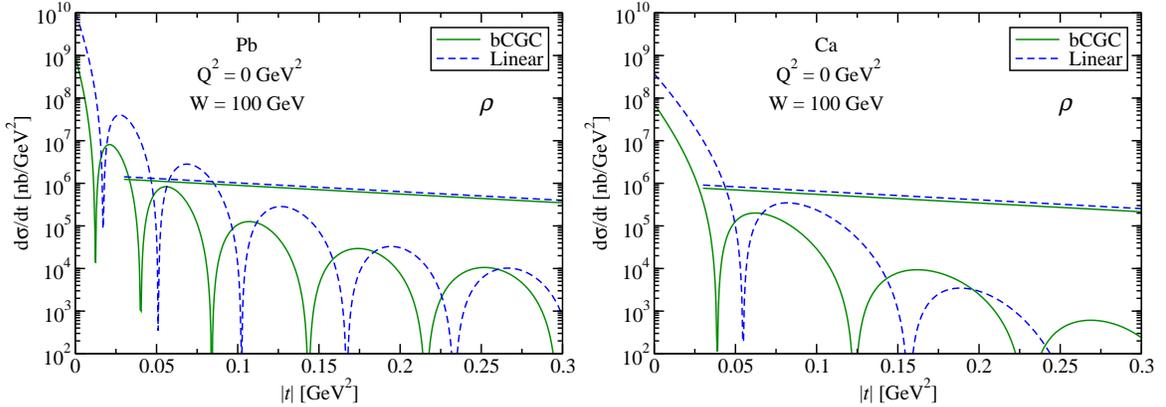

\begin{center}
 \includegraphics[scale=0.3]{dsdt_rho_pb_lin.eps}
 \includegraphics[scale=0.3]{dsdt_rho_ca_lin.eps}
\caption{(Color online) Comparison between non-linear and linear predictions for  the differential cross sections in coherent  and incoherent interactions. Predictions obtained using the  Gaus-LC  model for the $\rho$ wave function.}
\label{dsdt_din}
\end{center}
\end{figure}

\begin{figure}[t]
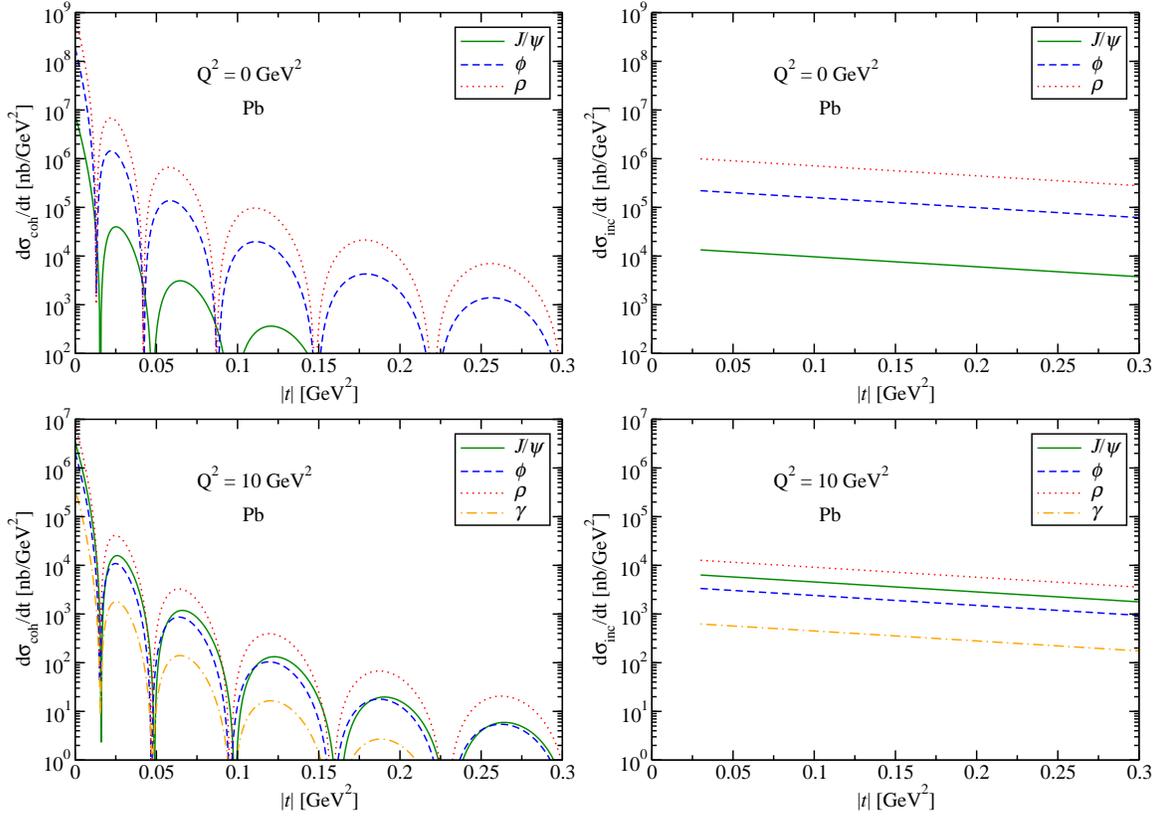

\begin{center}
 \includegraphics[scale=0.3]{coh_dsdt_vm_q0.eps}
 \includegraphics[scale=0.3]{inc_dsdt_vm_q0.eps}\\
 \includegraphics[scale=0.3]{coh_dsdt_vm_q10.eps}
 \includegraphics[scale=0.3]{inc_dsdt_vm_q10.eps}
\caption{(Color online) Comparison between the differential cross sections for coherent  and incoherent  production of different final states considering 
$Q^2 = 0$ (upper panels) and $Q^2 = 10$ GeV$^2$ (lower panels). Predictions obtained using the  Gaus-LC and bCGC models.}
\label{dsdt_comp}
\end{center}
\end{figure}

\section{Results}
\label{res}

In what follows we will present our predictions for the coherent and incoherent $\rho$ production considering different models for the dipole - proton scattering amplitude and for the $\rho$ wave function. Our main focus will be in the kinematical region of small values of the Bjorken - $x$ variable that will be probed in the future electron - ion collider facility to be constructed in the USA \cite{Accardi}. Basically, we will restrict our analysis to small virtualities ($Q^2 \le 10$ GeV$^2$) and center - of - mass energies $W$ of the order of 100 GeV.     However, it is important to emphasize that the magnitude of the non-linear effects increases with the energy, which implies that our main conclusions should also be valid for the diffractive $\rho$  production in $eA$ collisions at the LHeC \cite{LHeC}.

Initially let us  address the energy $W$ and virtuality $Q^2$   dependencies of the coherent and incoherent cross sections considering  two different 
nuclei, $A = 40$ (Ca) and 208 (Pb), and the Gaus-LC model for the overlap function. Moreover, two distinct models for the dipole - proton scattering amplitude are used to estimate the cross sections.
In the left panels of the Fig.~\ref{sec_energy} we present the energy dependence of our predictions assuming $Q^2 = 0$. We observe that the cross sections increase with the energy and nuclear mass number. The coherent processes have a stronger $A$ dependence in comparison to the incoherent one, which implies that the dominance of the coherence processes increases with the atomic number. Regarding the dependence on the model used for the dipole - proton scattering amplitude, we have that
the  bCGC and rcBK models yield very similar predictions at large nuclei. At lighter nuclei the bCGC predictions are $\approx 15$ \% larger than the rcBK one, which is directly associated to the  different transition between the linear and non-linear regime predicted by these models (See, e.g. Fig. 3 in Ref. \cite{diego}). 
This distinct behavior has direct impact on the 
$Q^2$ dependence of the coherent and incoherent cross sections, presented in the right panels of   Fig.~\ref{sec_energy}. We have that the cross sections decrease with the photon virtuality, with the difference between the bCGC and rcBK predictions increasing with $Q^2$. Consequently, the probe of the cross sections at intermediate $Q^2$, e.g. $Q^2 = 10$ GeV$^2$, can be useful to constrain the description of the dipole - proton scattering amplitude.

As discussed in the previous Section and shown by the HERA data, the diffractive $\rho$ production  probes the transition  between the soft and hard QCD regimes, with the energy dependence being strongly sensitive to $Q^2$. In particular, at small - $Q^2$, the energy dependence of the $ep$ cross section is similar to that expected for soft processes ($\sigma \propto W^{\delta}$, with $\delta \approx 0.22$), while for large $Q^2$ one have $\delta \approx 0.8$, compatible with the dependence expected for hard processes.  In order to check how this behavior is modified by the nuclear medium, 
in Fig. \ref{slopes} we present our predictions for the energy dependence of the coherent (upper panels)  and incoherent (lower panels) cross sections considering different values of the photon virtuality. In order to facilitate  the comparison,  the distinct predictions for   $W = 50$ GeV have being  normalized to the unity. Moreover, for completeness, in the left panels we present the predictions for the $ep$ cross sections. As in the proton case, the slope of the distinct curves increases with the virtuality, with the growth  being smaller for heavier nuclei. Moreover, we 
can observe that the energy dependence is strongly modified at larger nuclei and smaller $Q^2$.
 This behaviour is expected, since in this kinematical 
range the magnitude of the non-linear effects is predicted to be amplified. At large $Q^2$, the difference between the slopes for different atomic mass number decreases, which is associated to the fact that the linear regime becomes dominant. Therefore, the study of the diffractive $\rho$ production at large nuclei and small - $Q^2$ is ideal to probe the gluon saturation effects in the QCD dynamics.

Let us now extend our analysis to the differential distributions $d\sigma/dt$ for coherent and incoherent interactions. In the case of the incoherent predictions we only present predictions for $|t| \ge 0.03$ GeV$^2$, since the model proposed in Ref. \cite{Lappi_inc} and used in our calculations fails to describe the vanishing of the incoherent cross section as $|t| \rightarrow 0$. Our motivation to analyse in detail these processes is associated to the fact that the precise separation of these events will allow us to access the spatial distribution of the gluon density in the nucleus through the  Fourier transform of the coherent cross 
section \cite{Toll} as well as to  study the incoherent processes which probe the  fluctuations in the interaction strengths of parton configurations in the nuclear wave function  with large values of the saturation scale \cite{Lappi_bal}. 
 In Fig. \ref{dsdt0} we 
present our predictions for the coherent and incoherent $\rho$ cross section for different nuclei and $Q^2 = 0$ considering the bCGC and Gaus-LC models for the dipole - proton amplitude and $\rho$ wave function. We find that the coherent cross section clearly exhibits the typical diffractive pattern, with  the dips in the range $|t| \le 0.3$ GeV$^2$  
increasing with the mass atomic number. Moreover, the coherent processes are characterized by a sharp forward diffraction peak and the incoherent one by a $t$ - dependence very similar to that for the $\rho$ production off free nucleons. The incoherent processes dominate at large - $|t|$  and the coherent ones at small values of the momentum transfer. This behaviour can be easily understood: with the increasing of the momentum kick given to the nucleus the probability that  it  breaks up becomes larger. As a consequence, the $eA$ interactions at large - $|t|$ are dominated by incoherent processes. Therefore, the analysis of the $t$  dependence can be useful to separate coherent and incoherent interactions. However, as discussed in detail in Refs. \cite{Caldwell,Toll}, the experimental separation of these processes is 
still  a challenge.  An alternative is the detection of the fragments of the nuclear breakup present in the incoherent processes. e.g. the detection of emitted neutrons by zero - degree calorimeters.

In Fig. \ref{dsdt1} we analyse the dependence of our predictions on the basic inputs of our calculations.  In particular, in the left panels we present our predictions considering different models for the dipole - proton scattering amplitude and in the right panels for different wave functions. We find that the positions of the dips are almost independent of the dipole - proton model used as input in the calculations, as already observed in Ref. \cite{diego} for the nuclear DVCS. Moreover, we obtain that our results are almost independent of the model used to calculate the wave functions. This conclusion is also valid for the incoherent cross section, which is slightly dependent on the model used for ${\cal N}^p (x, \rr)$ for lighter nuclei.

In Fig. \ref{dsdt2} we analyse the dependence on $Q^2$ of our predictions considering the bCGC and Gaus-LC models for the dipole - proton scattering and $\rho$ wave function, respectively. Moreover, we present predictions for $A = Pb$ (left panels) and $A = Ca$ (right panels).  As expected from the analysis of the total cross sections, the differential  distributions decrease with the photon virtuality. Our results demonstrate that the position of the dips in the coherent cross sections 
are almost independent of $Q^2$ and that the $t$-dependence of the incoherent cross sections are similar for different values of the photon virtuality.

One important aspect to be investigated is the impact of the non-linear effects in the diffractive $\rho$ production. As discussed above, the coherent and incoherent cross sections are determined by the dipole - nucleus $\N^A$ and dipole - proton  $\mathcal{N}^p$ scattering amplitudes, respectively.
In order to estimate the magnitude of the non-linear effects, let us assume that dipole - nucleus amplitude can be expressed  by
\begin{eqnarray}
  \N^A(x,r,b) =  \frac{1}{2}\sigma_{dp}(x,r)AT_A(b)
  \label{nalin}
\end{eqnarray}
with $\sigma_{dp}$ expressed by Eq. (\ref{sdip}). Moreover, we will assume in the calculation of $\sigma_{dp}$ and  the incoherent cross section that
 $\mathcal{N}^p(x,\rr,\rb)$ is given by the linear part of the bCGC model, which is 
\begin{widetext}
\begin{eqnarray}
\mathcal{N}^p(x,\rr,\rb) =  
{\mathcal N}_0\, \left(\frac{ r \, Q_s(b)}{2}\right)^{2\left(\gamma_s + 
\frac{\ln (2/r Q_s(b))}{\kappa \,\lambda \,Y}\right)}\,\,,
\label{eq:bcgclin}
\end{eqnarray}
\end{widetext}
with the same parameters used before in Eq. (\ref{eq:bcgc}). The basic idea in Eq. (\ref{nalin}) is to disregard the multiple scatterings of the dipole with the nuclei, which are taken into account the non-linear effects in the full calculation. On the other hand, Eq. (\ref{eq:bcgclin}) means that we are disregarding  possible  non-linear effects in the nucleon.
In Fig. \ref{dsdt_din} we present a comparison between linear and non-linear predictions.
We can observe that the incoherent cross sections are not strongly modified by the non-linear effects. In contrast, in the case of
 coherent interactions, the magnitude of the cross section and position of the dips are distinct in the non linear and linear predictions, 
which makes the analysis of this observable a sensitive probe of the non-linear QCD dynamics.

Finally, let us compare our predictions for the $\rho$ production with those for other exclusive final states: $J/\Psi, \, \phi$ and $\gamma$. Our motivation to perform this comparison is associated to the  fact that the lighter vector mesons ($\rho$ and $\phi$) are more sensitive to saturation effects than the $J/\Psi$, since 
the corresponding  overlap functions peak at larger pair separations at a fixed photon virtuality $Q^2$. Moreover, this study allows to  probe the transition between the non-linear and linear regimes of the QCD dynamics by increasing the photon virtuality. 
In Fig. \ref{dsdt_comp} we present our results for two different values of $Q^2$ and $A = Pb$. For $Q^2 = 0$ (upper panels) we observe that the differential cross sections decrease at heavier vector mesons and the position of the dips for the $J/\Psi$ production is slightly 
different of the other mesons, which is associated to the dominance of small dipoles in this final state. In contrast, $Q^2 = 10$ GeV$^2$ (lower panels), the position of the dips becomes almost identical for all exclusive final states. Therefore, the experimental analysis of heavy and light mesons at small - $Q^2$ can be useful to probe different regimes of the QCD dynamics.

\section{Conclusions}
\label{conc}

The study of exclusive processes in deep inelastic scattering (DIS) processes  has been one of the main topics in hadronic physics during the 
last years. It is expected that the  analysis of these processes allow us to  probe  the QCD dynamics at high 
energies,  driven by the gluon content of the target (proton or nucleus) 
which is strongly subject to non-linear effects (parton saturation). In particular, in electron - nucleus collisions the gluon density  is amplified due to the coherent contributions from many nucleons. Our goal in this
 paper was to extend and complement previous studies about the vector meson production in $eA$ collisions, presenting a comprehensive analysis
 of the energy, virtuality, nuclear mass number and transverse momentum dependencies of the cross sections for the $\rho$ production in the 
kinematical range which could be accessed in future electron - ion colliders.  We have found that the energy dependence of the differential
 cross sections are strongly modified with the increasing of the nuclear mass number and that coherent cross section dominates at small $t$ 
and the incoherent one dominates at large values of $t$. Our results demonstrated that the number of dips at small $t$ increases with the atomic number, with the 
position of the dips being almost   independent of the model used to treat the dipole - proton interaction.  
Moreover, we show that our predictions 
are insensitive to the model used to treat the $\rho$ wave function. A comparison with other exclusive final states was presented and we have 
demonstrated that the non-linear effects change the  positions of  the dips with respect to the linear regime. These results are robust 
predictions of the saturation physics, which can be used to investigate non-linear QCD dynamics in the kinematical range of 
future electron - ion experiments.


\begin{acknowledgements}
VPG would like to thank S. Klein and H.~Mantysaari  by useful discussions about  incoherent processes. This work was  partially financed by the Brazilian funding agencies CNPq, CAPES and FAPERGS.
\end{acknowledgements}

\end{document}